# The Payload Data Handling Unit (PDHU) on-board the HERMES-TP and HERMES-SP CubeSat Missions

A. Guzman*[a], S. Pliego[a], J. Bayer[a], Y. Evangelista[b], G. La Rosa[b], G. Sottile[b], S. Curzel[c], R. Campana[b], F. Fiore[b], F. Fuschino[b], A. Colagrossi[d], M. Fiorito[c], P. Nogara[b], R. Piazzolla[b], F. Russo[b], A. Santangelo[a], C. Tenzer[a]

[a]Eberhard Karls Universität Tübingen, Institute for Astronomy and Astrophysics, Tübingen, Germany; Istituto Nationale di Astrofisica-INAF, Roma, Italy; [c] Dipartimento di Elettronica, Informazione e Bioingegneria, Politecnico di Milano, Italy; [d] Dipartimento di Scienza e Tecnologia Aerospaziali, Politecnico di Milano, Italy

## ABSTRACT

The High Energy Rapid Modular Ensemble of Satellites (HERMES) Technological and Scientific pathfinder is a space borne mission based on a constellation of LEO nanosatellites. The payloads of these CubeSats consist of miniaturized detectors designed for bright high-energy transients such as Gamma-Ray Bursts (GRBs). This platform aims to impact Gamma Ray Burst (GRB) science and enhance the detection of Gravitational Wave (GW) electromagnetic counterparts. This goal will be achieved with a field of view of several steradians, arcmin precision and state of the art timing accuracy. The localization performance for the whole constellation is proportional to the number of components and inversely proportional to the average baseline between them, and therefore is expected to increase as more. In this paper we describe the Payload Data Handling Unit (PDHU) for the HERMES-TP and HERMES SP mission. The PDHU is the main interface between the payload and the satellite bus. The PDHU is also in charge of the on-board control and monitoring of the scintillating crystal detectors. We will explain the TM/TC design and the distinct modes of operation. We also discuss the on-board data processing carried out by the PDHU and its impact on the output data of the detector.

## 1. INTRODUCTION

Our understanding of highly energetic transients of astrophysical origin has had great advances in the last decades [1] . Particularly, Gamma-Ray Bursts (GRBs) are one of the most intriguing and challenging phenomena for modern science. The High Energy Rapid Modular Ensemble of Satellites (HERMES) Technological and Scientific pathfinders aim to develop an all sky monitor looking for transients in the energy range between a 3-4 keV to 2 MeV[2]. Using a state-of the art timing resolution and baseline distance of thousands of km, this constellation of LEO nanosatellites will procure a fast (a matter of tenths of minutes or less) and accurate (arcmin) localization for these phenomena. These reduced localization times, as was the case in the discovery of Gravitational Waves and their electromagnetic counterparts, will allow a follow up in different wavelengths contributing to the multi-messenger astronomy. The scientific bounty of this observations is vast and includes the study of physics of matter in extreme condition, black holes, fundamental physics and the mechanisms of gravitational wave signal production which are associated with peculiar core-collapse supernovae and with neutron star/black hole mergers. More details on the scientific goals from the HERMES mission are provided in [2][3].

The HERMES project builds upon the present nanosatellite technologies that offer solid readiness at a limited cost. The philosophy is to use distributed modular components, instead of the traditional more expensive approach of flying a larger size telescope (hundredths of kilograms). Conversely, the HERMES detector's physical size must fit within a nanosatellite unit structure ($10 \times 10 \times 10$ cm$^3$) with a weight in the order of 1 kg. Due to these constraints, the effective area of each element is the order of 50 cm$^2$. This limitation is counteracted by the high multiplicity of single elements. The goal is to fly a constellation with a hundredth or so such units. This will achieve a whole sky monitor with an effective area of ~ 1 m$^2$. To allow for quick follow up of the potential candidates, each single unit must work with great sensitivity and provide a fast on-board pre-processing of the data. The Payload Data Handling Unit (PDHU) of the HERMES mission is in charge of controlling the detector and providing these pre-processing capabilities.

*guzman@astro.uni-tuebingen.com; phone 49 7071 29-75279; fax 49 7071 29-3458

In the present article, we commence by presenting a brief summary of the design of the HERMES payloads. We follow by a more detailed description of the PDHU hardware & software as well as their capabilities. And finally, we conclude presenting the hardware thus developed for the HERMES mission and the on-going pre-flight tests.

## 2. HERMES PAYLOAD DESCRIPTION

The current design of the HERMES payload is allocated in a 1U-Cubesat ($10 \times 10 \times 10$ cm$^3$). A exploded-out diagram is shown in Figure 1. We distinguish three main components: detector core, readout electronics, the power supply unit and the data handling unit. The core of the detector constitutes an array of 60 scintillator pixels optically insulated from each other but optically coupled to an array of 120 Silicon Drift Detectors (SDD). This configuration allows to detect both soft X-rays (directly absorbed in the SDD) and the scintillation light emitted by the crystals after the interaction with a higher energy photon. Each scintillator crystal is couped to two SDDs. Therefore, the scintillator light illuminates two SDDs, which permits a straightforward discrimination by the SDD-multiplicity of the events. In this scheme, scintillator events (double SDD) events correspond to γ-rays and single SDD events correspond to X-rays. The energy range of this configuration is between 3-5 keV up to 2 MeV[2].

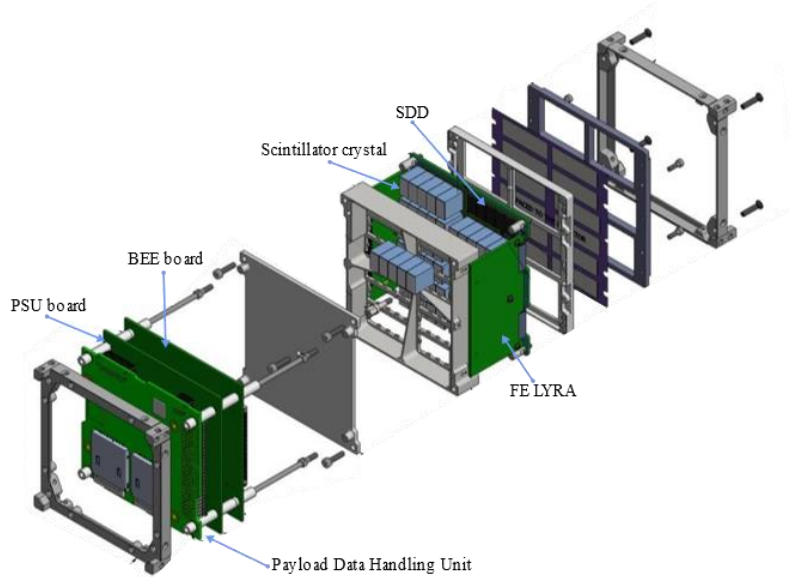

Figure 1 A schematic depiction of the HERMES payload. Modified from image courtesy of R. Piazzolla.

The readout electronics designed and developed for the HERMES detectors are named LYRA Application Specific Integrated Circuit (ASIC). The LYRA-ASICs have heritage from the VEGA-ASICs developed for the LOFT Phase A study[4]. Each single LYRA-ASIC operates as a set of 32+1 Integrated Circuits (ICs). This ICs are grouped in 32 Front End ICs (FE-LYRA) per one Back End IC (BE-LYRA). The FE-LYRAs pre-amplify and do the initial shaping of the signal coming from the SDDs. The BE-LYRA takes these 32 inputs and finalizes the signal processing. The FE-LYRA ICs are placed very close to the SDD nodes in order to minimize stray capacitances, whereas the BE-LYRAs are placed on the Back End Electronics (BEE) board.

Additionally, the BEE allocates external analog to digital converters and an FPGA in charge of the control logic. This FPGA reads out the analogue signals of the BE-LYRA ICs and synchronizes the digital conversion operations and the time tagging of the events. The time tagging combines a conventional GPS timing (based on a commercial GPS device) and the novel implementation of Chip Scale Atomic Clock (CSAC) to reduce the natural jitter of the GPS sensor to ensure sub-microsecond timing resolution. The BEE is also in charge of automatically discriminating the multiplicity of the readout signals. This task is crucial to generate a real-time photon list that includes a preliminary energy assignment to the detected photons as well as the timing information.

Finally, before continuing to the next subsystem and the main topic of this article, we mention the Power Supply Unit (PSU). The PSU is the payload's component connected to the satellite power bus, in charge of the control and generation of the rest of payload power supplies.

## 3. THE PAYLOAD DATA HANDLING UNIT (PDHU)

The PDHU oversees and controls the payload's operations; pre-processes, stores and prepares the scientific data; logs the payload's housekeeping, and serves as the interface between the rest of the payload subsystems and the service module (also referred in this text as the satellite bus).

The selected hardware for the PDHU is the Innovative Solutions In Space (ISIS) On-Board computer (iOBC)[5]. The iOBC is a high performance & low power processing unit based around a 400 MHz 32bit ARM9 processor, and with multiple standard interfaces. The iOBC was designed specifically for use in Nano-Satellites and has flight heritage since 2012. Particularly important for the HERMES mission is the low power consumption (around 400 mW), the redundant 2 GB non-volatile data storage (with industrial grade SD-Cards) and the 512 kB FRAM which for practical purposes is impervious to single event upsets.

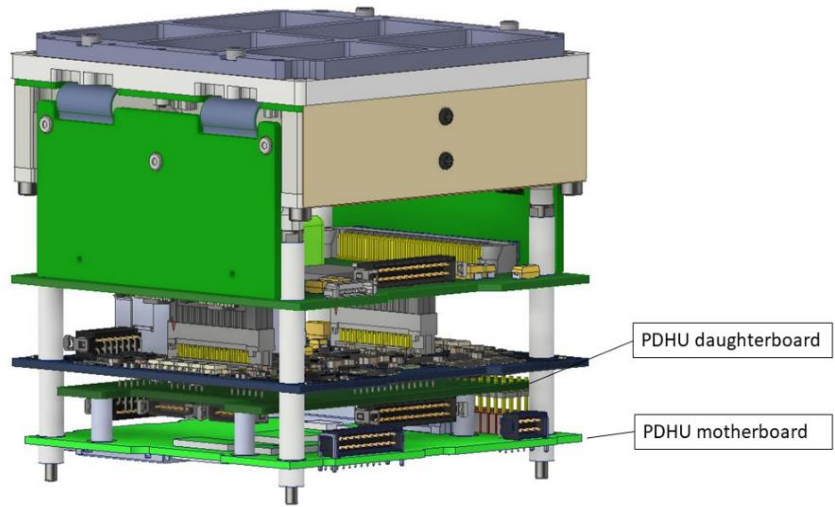

Figure 2: The position of the PDHU within the payload. The PDHU daughterboard is explained below in the text. Image courtesy of R. Piazzolla.

**The PDHU Daughter board**

Part of the advantages of using Consumer Off The Shelf (COTS) products is the adaptability gained by using readily available technological solutions. To interface the iOBC with different peripherals/subsystems, the iOBC offers the possibility to adapt a custom-made daughterboard. We developed the PDHU-daughterboard (DB) to match iOBC with the HERMES requirements. The DB consists of a shield-design daughterboard which interconnects the iOBC with the rest of the payloads subsystems. PC-104 connectors are commonly used on the CubeSat standard. However, HERMES module is not adopting this connection philosophy. Therefore, the daughter board routes the main power and basic communication signals through a couple of slimmer SAMTEC MTLW-126-07-G-D-265 connectors (see Figure 3). As a side effect the iOBC's manufacturer had to customize their design by replacing the standard PC-104 connector with two SAMTEC SLW-126-01-G-D connectors.

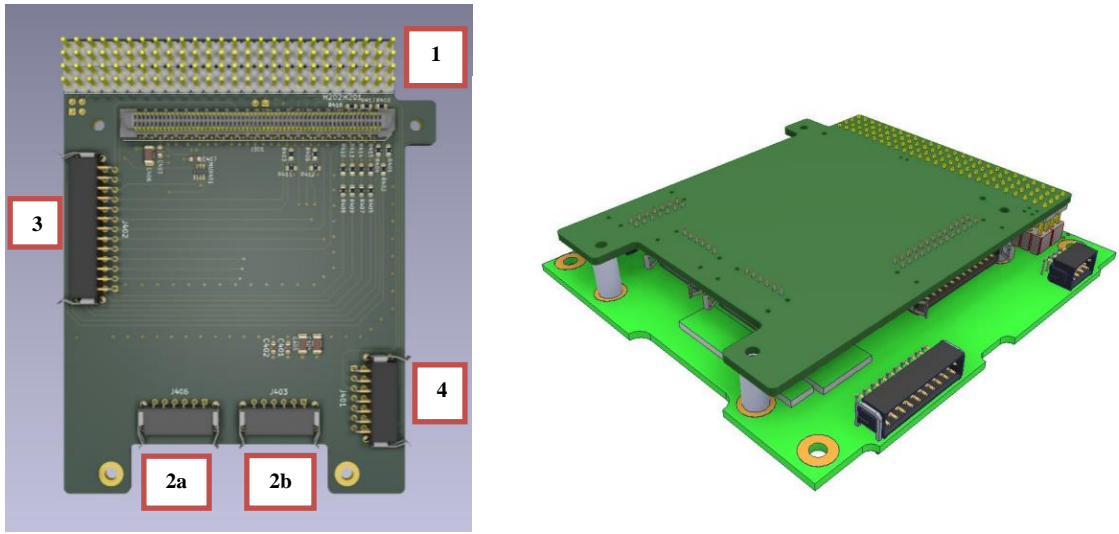

Figure 3: A CAD model of the PDHU's custom-designed DB. On the left side we show the bottom view. The connectors are numerated 1) the PC-104 "replacement" connector. 2a & 2b) the connectors to the satellite bus. 3) the BEE connector. 4) the PSU connector. To the right side iOBC and DB assembled together (top view).

### 3.1 Interfaces

As mentioned above the PDHU is the main interfaces between the payload and the satellite bus. In this subsection we discuss the characteristics of each interface. These are conceptually depicted in Figure 4.

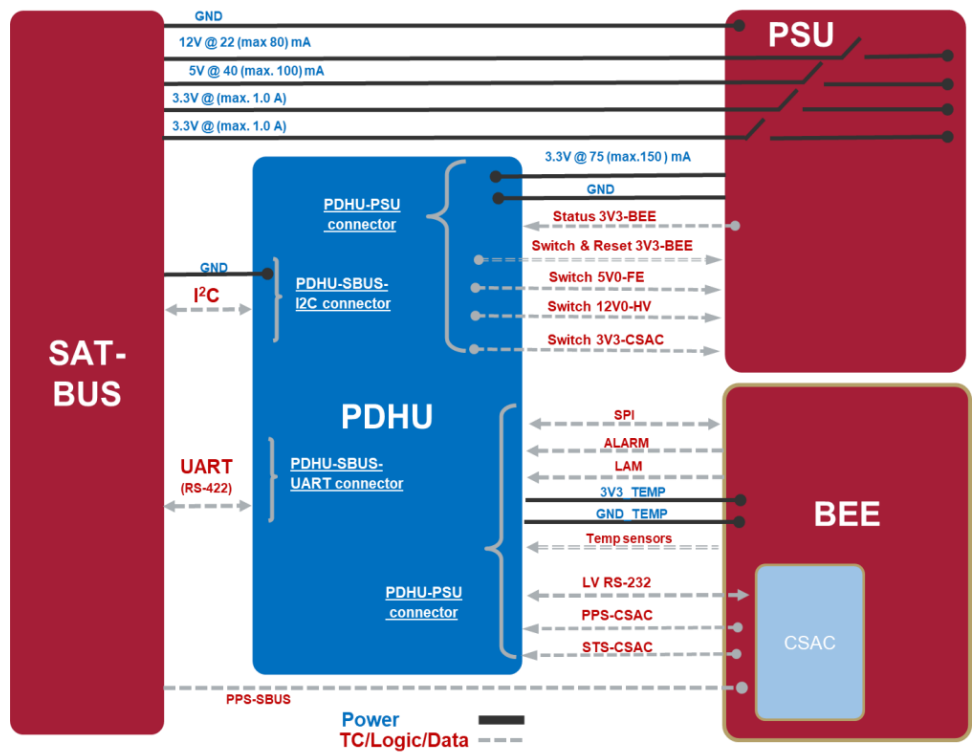

Figure 4. The interfaces between the PDHU (blue boxes) and other subsystems (burgundy). The black solid lines indicate voltage lines (and switches), while the grey dashed lines indicate communication interfaces. Note that the CSAC is integrated Into the BEE electronics.

**SBUS interface:**

There are two connectors shown on Figure 3 contain the communication channels between the SBUS and the PDHU. One of the connectors carries the signals for an I$^2$C bus with the SBUS as master, and the other one a serial bus (UART) over RS422 standard. Both communication channels are redundant. However, due to the difference in transmission speeds, the I$^2$C is preferable solely used for TC and to read some HK data. On the other hand, scientific data and more elaborated status reports are transmitted preferably via the serial bus. After any issued telecommand the PDHU emits an "acknowledge"-response nominally in the same channel where the TC was issued. However, this latter channel can be pre-defined to be either the I$^2$C or the UART. Also present in the connector of the I2C bus is a single UART line (following the LV-RS232 standard) transmitting the GPS information directly from GPS receiver on board the service module.

**PSU interface:**

The power supply lines (GND and 3.3 V) for the PDHU arrive from the PSU. There are 4 lines departing from the PDHU to operate 4 voltage switches on the PSU. These switches are operated by the PDHU but placed on the PSU. They are switches for the 3.3 V, 5 V and 12 V lines that power the detector; plus an extra 3.3 V line that powers the BEE. A digital reset line for the 3.3 V BEE and a digital line codifying the status of the 3.3 V BEE line are also present and GPIO pin signaling

**BEE interface:**

The main communication channel between BEE and PDHU entails an SPI-bus with the PDHU as master. Additionally, there are two digital lines to allow the BEE to signal data ready i.e., Look-At-Me (LAM) signal and a House Keeping (HK) alarm. In addition, seven analog lines coming from seven temperature sensors are also read. Also present in the BEE is the CSAC, which the PDHU controls using a serial connection (UART). The PPS signal produced by the CSAC is also connected to the PDHU.

### 3.2 Operating modes

The PDHU software design follows a finite-state-machine model that simplifies the operation of the HERMES payload. The different operating modes or states are depicted in Figure 5. Essentially all operating modes are one step away from the standby mode, the latter being the default mode after boot.

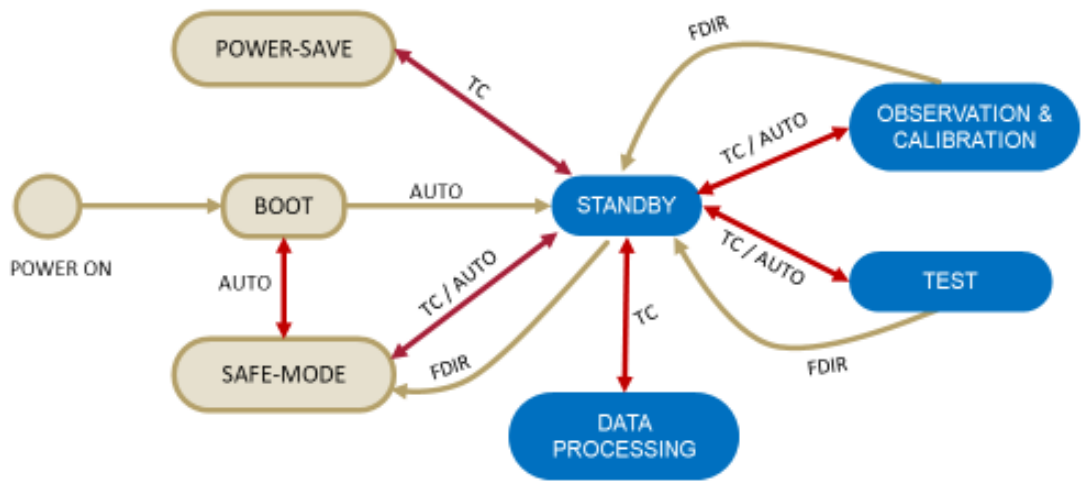

Figure 5. A diagram illustrating the finite-state model of the PDHU's operating modes.

**BOOT**

It is the start-up mode at power-on. This loads the processor code from the flash memory of the iOBC. The bootloader and necessary initializations routines are procured by the iOBC's manufacturer. Nominally this mode moves automatically to standby, unless there is an un-attended failure in the system (overcurrent, temperature outside valid range etc.)

**SAFE-MODE**

In this mode only the PDHU has very limited functionality (mostly only TC and some diagnostics). It is triggered via either TC or when a major error/malfunction has been detected.

**STANDBY**

After power-on the PDHU moves to STANDBY mode to start the instrument monitoring and control. Nominally, before and after observations this is the "go-to" operational mode. However, as part of the design, this mode can be reached via the Fault Detection, Isolation and Recovery (FDIR) procedures. Our FDIR can cope with simple retrying or power cycling of certain components, while logging the faults. This log as well as the result of the alleviating procedures is transmitted to the main On-Board Data Handling computer (OBDH) on the service module. Housekeeping routines are running. All other operating modes are reachable after the suitable TC.

**OBSERVATION AND/OR CALIBRATION**

Strictly speaking this are too distinct operating modes, specially from the detector point of view where specific calibration procedures are foreseen. However, from the point of view of the PDHU, both instrument-calibration and scientific-acquisition modes resemble a general OBSERVATION mode whose particular details are configured while in STANDBY mode. Some event pre-processing is foreseen in this mode (see burst search below), although the priority is the correct control and data readout of the detectors. It is foreseen to initiate this mode with a TC or on a scheduled basis.

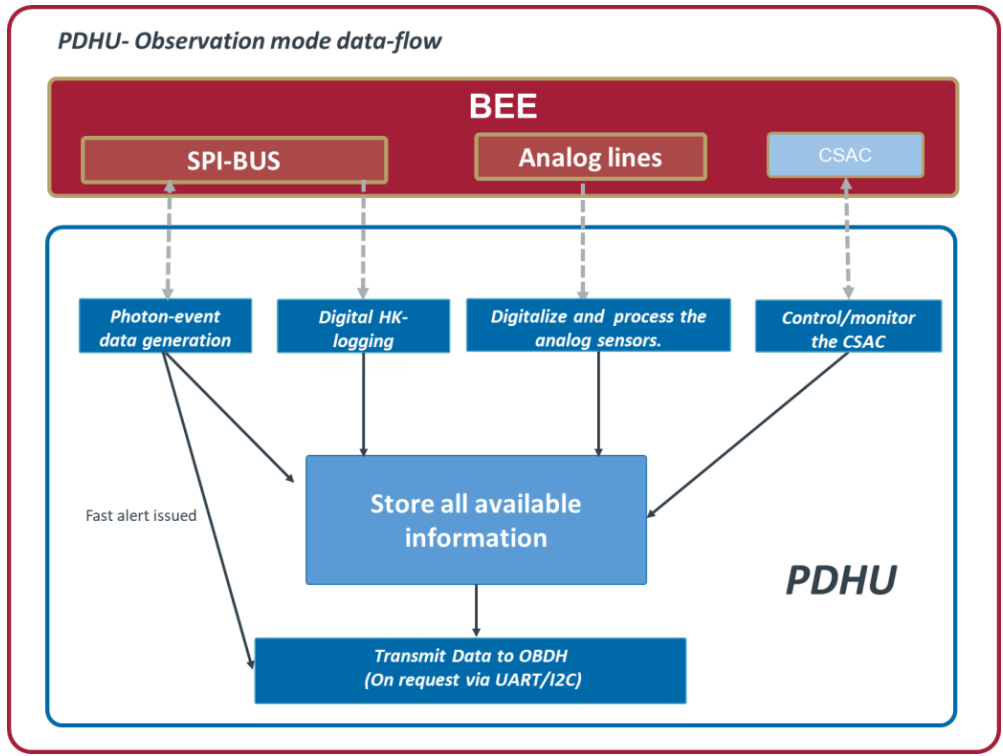

Figure 6. Parallel tasks running on the PDHU during OBSERVATION mode.

The different tasks run by the PDHU during OBSERVATION mode are sketched in Figure 6. These processes are run in parallel using suitable scheduling algorithms provided by the Real Time Operating System (RTOS[6], that is running on the iOBC. The general tasks are:

- Photon event data generation: Receive and pre-process the photon-by-photon information acquired by the detector. Provides a real-time "burst alarm" on received data (see below).

- Digital HK-logging. Handle the digital elements of the housekeeping values reported by the BEE (voltages and currents amongst others). Monitors that these values are within prescribed limits and logs them for possible downlink.

- Digitalize and process analog sensors. Use the iOBC's eight 10 bit ADCs to digitize and record the the "stand alone" analog sensors (temperature and voltages) present on the payload.

- Control monitor the CSAC: During observation the correct functioning of the CSAC is monitored by the PDHU.

**TEST**

This mode is introduced to implement a merge of STANDBY and OBSERVATION modes. Its main purpose is to perform specific checks for diagnosis purposes. This mode is also only to be triggered via TC. Some of the foreseen tests include (the following list is not exhaustive): ADC-readout test, SPI-communications test, BEE readout test, and CSAC test.

**DATA PROCESSING**

Its main purpose is to prepare the scientific data packets for transmission to ground. This includes on-board burst-searching algorithms (see below) and data reducing/compressing. Although this operational mode is conceived as a stand-alone mode, running some of its tasks while in OBSERVATION mode is also a possibility. This would allow for a prompter acknowledgement of scientifically interesting events. Nominally this mode is the go-to mode after a given observation is finished.

**POWER SAVE**

This mode is introduced to implement a low power consumption mode. Its main purpose is to lower the HERMES payload power consumption to cope with time periods of reduced power availability from the spacecraft bus. The POWER SAVE mode reduces the running processes on the PDHU to a minimum as well as turning off most of the payload component while monitoring the payload's health status. This mode is triggered via either TC or a time-scheduled change in the operation mode.

## 4. ON-BOARD DATA HANDLING

The data arriving to the PDHU needs to be processed (reduced) before being transmitted to ground. Simultaneously, it is of paramount importance to detect a transient event in the relevant energy ranges as early as possible to allow for a follow up observation. This procedure commences by differentiating between events. For X-ray events only one amplitude of the two per ASIC channel contains useful information. For $\gamma$ events the energy is deposited in the scintillator crystal and therefore both amplitudes of the ASIC channel contain useful information. The number of triggered ASICS is reported by the BEE and readout by the PDHU. Multiple X and/or $\gamma$ event will be recognized following the same logic of the two previous cases. They will share the same time stamp, this temporal check is carried out by the PDHU. This information and housekeeping logs are saved by the PDHU and transmitted upon request to the OBDH. This concept is illustrated in Figure 7.

Due to data downlink constraints, a full record of photon events detected by the payload can be downlinked only for transient phenomena (GRBs, solar flares or other fast transients). To perform those functions, the following components in the PDHU system are required:

- Payload Event Buffer (EB). This is essentially a complete list of all detected events that can be kept in the on-board mass memory. These buffers need to be further processed to get a list of 'cleaned' events;

- Detector Ratemeters (DRs) providing count-rates for different time scales from milliseconds to several seconds (e.g. 0.001, 0.016, 0.064, 0.256, 1, 10, 100 seconds). The DRs are separated also by energy range and geometrical region of the detector (quadrant).

- Burst Search routines for short timescales, e.g. 256 ms, 16 ms, 1 ms and sub-ms;

- Burst Alarm, to be raised and communicated to the SM in case a transient candidate has been found. The SM can then decide to transmit a prompt message indicating the satellite has detected a transient candidate and relay some information about the event (see *IRIDIUM* alerts below).

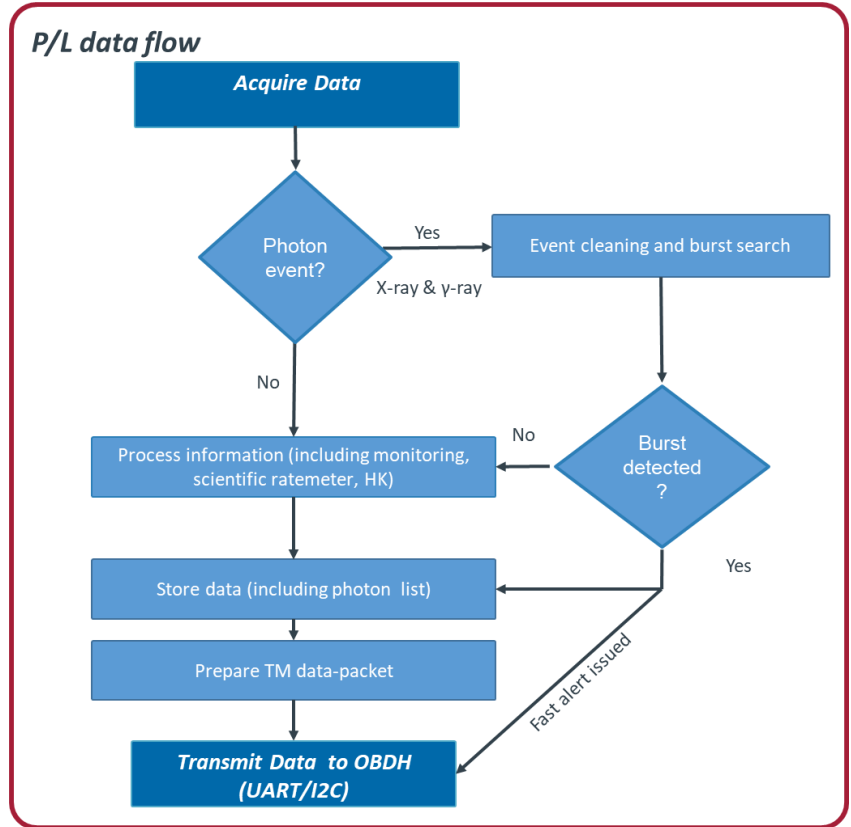

Figure 7. Block diagram of PDHU data type identification

### 4.1 Trigger strategy

The proposed triggering scheme is a multi-tier & multi-channel scheme. It includes the following levels:

**Level 1 triggers**

This trigger implemented at the BEE level. It is activated if the ADC-counts of a given channel are above an adjustable threshold. This threshold is configured via an ASIC-configuration table. This threshold is set with an 8 bit DAC assigned during calibration and configured via the ASIC-configuration table.

**Level 2 triggers or burst search algorithms**

Activated in the "real-time" (i.e. during OBSERVATION) at the PDHU level. The transient-search routine (referred to as "burst search" routine) finds a transient above an adjustable likelihood value. These triggers are applied at specific energy bands (three different energy bands) and for different time windows. An schematic of the procedure is shown in Figure 8.

This Burst Search (BS) algorithms rely on two approaches. The first approach is done by the PDHU using payload ratemeters. Based on different ratemeters integrating data in different energy bands, timescales and geometric regions.

This first approach determines the existence of a burst, defined as events producing count rates above thresholds determined themselves by the current background level. Since the GRBs are strongly energy and timescale dependent, we expect different results for the different ratemeters. The threshold energy defining the three energy channels are re-programmable on flight, as well as the integrating timescales. In a nutshell, the algorithm checks the condition shown in Eq. (1) for every ratemeter. Therefore, each ratemeter $R_i$ reflects both energy range (low, medium, high) labeled by the index i, and timescale τ. The idea illustrated in Eq. (1) is to check if any specific ratemeter $R_i$ is above a predefined static offset $O_i$, plus the scaled (via α) sum of the dynamically changing background $B_i$ and a number ($n$) of its standard deviations $\sigma_{Bi}$

$$R_i(\tau) > \alpha\big(B_i(\tau) + n\,\sigma_{B_i}\big) + O_i \qquad (1)$$

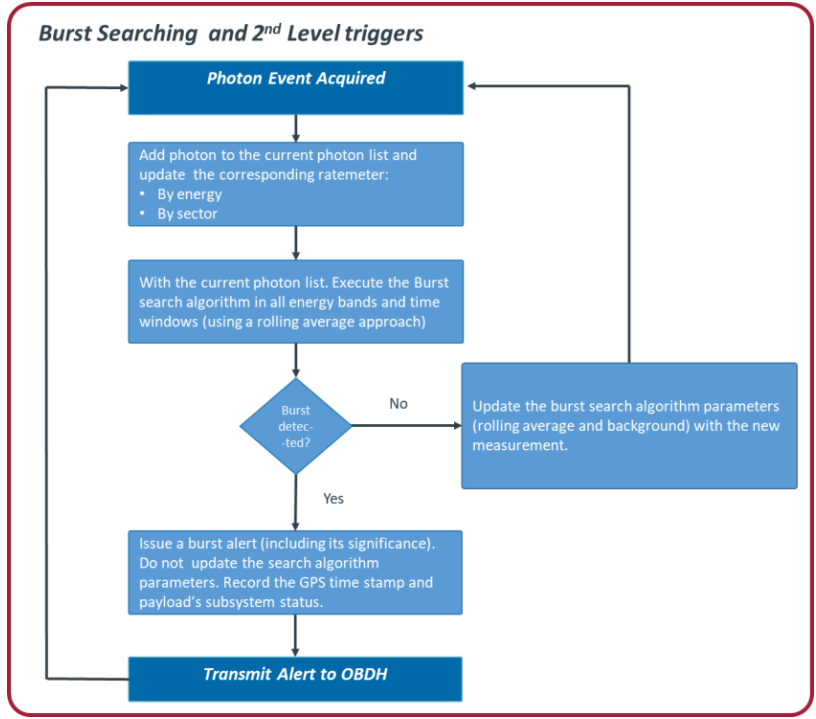

Figure 8. The burst searching algorithms running during observations.

The second approach for transient search follows a Bayesian approach [7] assuming a Poissonian distribution for the background. In this approach the instantaneous log-likelihood for after a series of measurements labeled *i* with $n_i$ number of counts respectively, during a time window *τ*, and with an expected counts $\lambda_i$, is computed as:

$$\ln L_{\tau,E} = \sum_i n_i^{\tau,E} \ln \lambda_i^{\tau,E} - \lambda_i^{\tau,E} - \ln n_i^{\tau,E}! \qquad (2)$$

In a similar way as the first burst alert, if the log-likelihood exceeds a set threshold given via TC, then a burst alert is issued.

**Level 3 triggers & Photon list cleaning procedure**

These triggers are applied during the data-process operating mode after the "event cleaning" routines have concluded (see below). These triggers also have their own set of adjustable parameters per energy band and time-resolution.

This procedure could take in account at least two different phenomena that could be mistaken as photon events: electronic retriggering and the passage of charged particles through the detector. The first phenom occurs when the same electronic channel retriggers when the corresponding SDD is hit by a particle with a large release of energy, saturating the amplifier with consequent "bouncing" the baseline near the threshold level. The cleaning procedure checks that within 100 μs the same SDD is not recorded more than N (e.g. 5) times. In case the check is positive the data from this SDD is discarded. Secondly, to discard charge particles posing as photon events. A detector wide search for saturated

SDDs that triggered on the same 100 ns window. Multiplicities above 3 are taken as evidence for a charged particle event

After the event cleaning procedure has finalized for a given photon list, similar transient-search algorithms as for level 2 triggers can run through the cleaned photon list in order to emit a burst-alert. The reader should keep in mind that a higher/lower trigger level does not necessarily correspond with a higher/lower significance of the event. They correspond with different time scales and different triggering algorithms (e.g. Level 3 are the ones that take longer to calculate either because of their complexity or due to longer acquisition times). An example of a simulation of the nominal behavior of the detector and the information processed by the PDHU is shown in Figure 9.

**IRIDIUM burst alerts**

Our transient event algorithm issues "Burst alerts" send to the service module. This potentially scientifically interesting observations are accompanied with "ready to use" data. They can be made available to the user much earlier than regular data as they would exploit the available telemetry budget with the IRIDIUM constellation. Such a Quick-Look Data (QLD) is not in the baseline data products for the HERMES mission since it is not required to fulfill the scientific objectives of the mission. However, it is highly desirable for increasing the scientific output of the mission, even if carried out on a best-effort basis. After a burst alert is issued the photon list will be searched to produce a 340 byte package to be transmitted immediately to ground containing the: satellite identification number, GPS information including location at the time of the triggered, three light curves at different energy bands containing the count-rates before, during and after the alert was issued, and ancillary data pertaining the health and overall status of the detectors (count rates, status of other triggers, temperatures, currents, etc.)

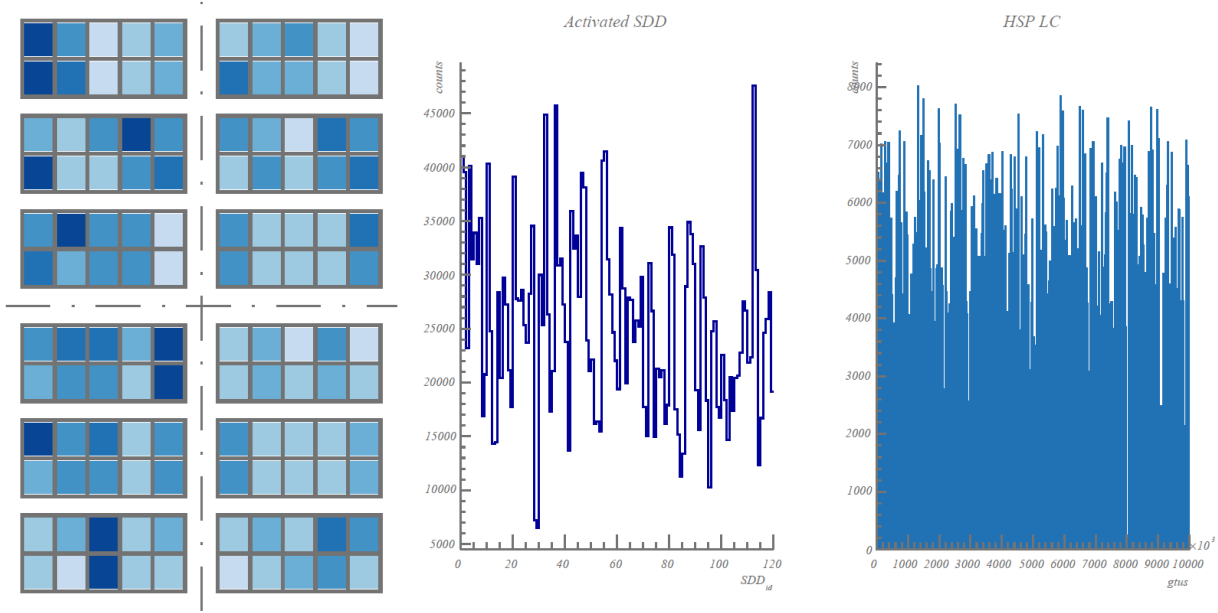

Figure 9. A computer simulated example of a one second acquisition using mock-up values for the count rates. To the left we see a "heat map" showing a schematic representation of the HERMES-detector's SDDs where a darker SDD corresponds to higher count rate. In the middle we see the the total counts for each individual SDD. And to the right we see the LC with a time resolution of 0.1 s labeled as gate-time-unit (gtu)

## 5. CONCLUSIONS

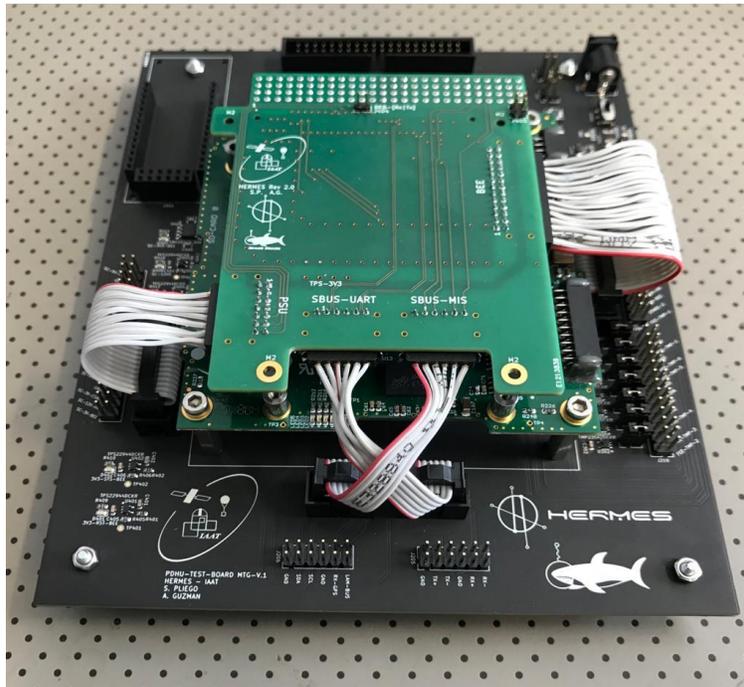

Figure 10. An engineering model of the PDHU sitting atop an custom designed debugging board,use to test the PDHU interfacing with other subsystems( in a *flat-sat* approach)

The authors have designed and implemented the PDHU as described in this article on engineering models of the flight hardware (see Figure 10). In addition to this, preliminary tests and simulations show the feasibility of what has been developed, to fulfill the requirements of the HERMES mission. In the coming months, a complete payload integration with the actual detectors is foreseen. The outcome of this integration will clarify and demonstrate the behavior of the payload as whole subsystem.

**Acknowledgments**

This work has been carried out in the framework of the HERMES-TP and HERMES-SP collaborations. We acknowledge support from the European Union Horizon 2020 Research and Innovation Framework Programme under grant agreement HERMES-Scientific Pathfinder n. 821896 and from ASI-INAF Accordo Attuativo HERMES Technologic Pathfinder n. 2018-10-H.1-2020.

## REFERENCES

[1] Frontera, F. et al. "The Gamma Ray Burst catalog obtained with the Gamma Ray Burst Monitor aboard BeppoSAX", The Astrophysical Journal Supplement Series, 180 (1) 192-223 (2008)
[2] Fiore, F. et al., "The HERMES-technologic and scientific pathfinder", Proc. SPIE 11444-166 (2020).
[3] Burderi, L. et al., "GrailQuest & HERMES: Hunting for Gravitational Wave Electromagnetic Counterparts and Probing Space-Time Quantum Foam", Proc. SPIE 11444-252 (2020).
[4] Evangelista, Y., et al., "The scientific payload on-board the HERMES-TP and HERMES-SP CubeSat missions", Proc. SPIE 11444-168 (2020).
[5] Innovative Solutions In Space, "ISIS On Board Computer", https://www.isispace.nl/product/on-board-computer/ (20 November 2020).
[6] Real Time Engineers Ltd, "FreeRTOS[TM] Real-time operating system for microcontrollers" https://www.freertos.org/index.html (20 November2020).
[7] Bélanger, G., "On detecting transient phenomena", The Astrophysical Journal, 773:66 (12pp) (2013).